\documentclass[runningheads]{llncs}

\usepackage{amssymb,amsmath,graphicx,times,cite,hyperref}
\usepackage[lined,ruled,boxed]{algorithm2e}
\newcommand{\keywords}[1]{\par\addvspace\baselineskip
\noindent\keywordname\enspace\ignorespaces#1}

\begin{document}

\mainmatter

\title{Transfer Learning Enhanced Common Spatial Pattern Filtering for Brain Computer Interfaces (BCIs):\\ Overview and a New Approach}
\titlerunning{Transfer Learning Enhanced CSP Filtering for BCIs: Overview and a New Approach}

\author{He He \and Dongrui Wu}
\authorrunning{He He and Dongrui Wu}

\institute{School of automation, Huazhong University of Science and Technology, Wuhan, China\\
\email{hehe91@hust.edu.cn}, \email{drwu@hust.edu.cn}}
\maketitle

\begin{abstract}
The electroencephalogram (EEG) is the most widely used input for brain computer interfaces (BCIs), and common spatial pattern (CSP) is frequently used to spatially filter it to increase its signal-to-noise ratio. However, CSP is a supervised filter, which needs some subject-specific calibration data to design. This is time-consuming and not user-friendly. A promising approach for shortening or even completely eliminating this calibration session is transfer learning, which leverages relevant data or knowledge from other subjects or tasks. This paper reviews three existing approaches for incorporating transfer learning into CSP, and also proposes a new transfer learning enhanced CSP approach. Experiments on motor imagery classification demonstrate their effectiveness. Particularly, our proposed approach achieves the best performance when the number of target domain calibration samples is small.

\end{abstract}

\section{Introduction}

Brain computer interfaces (BCIs) \cite{Wolpaw2002,Lance2012} provide a direct communication pathway for a user to interact with a computer or external device by using his/her brain signals, which include electroencephalogram (EEG), magnetoencephalogram (MEG), functional magnetic resonance imaging (fMRI), functional near-infrared spectroscopy (fNIRS), electrocorticography (ECoG), and so on. EEG-based BCIs have attracted great attention because they have little risk (no need for surgery), are convenience to use, and offer high temporal resolution. They have been used for robotics, speller, games, and medical applications \cite{Nicolas-Alonso2012,Erp2012}.

However, there are still many challenges for wide-spread real-world applications of EEG-based BCIs \cite{Makeig2012,Lance2012}. One of them is related to the EEG signal quality. EEG signals can be easily contaminated by various artifacts and noise, including muscle movements, eye blinks, heartbeats, environmental electromagnetic fields, etc. Common approaches to clean EEG signals including time-domain filtering and spatial filtering. Common spatial pattern (CSP) filtering \cite{drwuSF2018,Blankertz2008,Ramoser2000,drwuRG2017} is one of the most popular and effective spatial filters for EEG to increase its signal-to-noise ratio.

CSP performs supervised filtering, which requires some subject-specific calibration data to design. This is time-consuming and not user-friendly. A promising approach for shortening or even completely eliminating this calibration session is transfer learning (TL) \cite{Pan2010}, which has already been extensively used to handle individual differences and non-stationarity in EEG-based BCI \cite{Jayaram2016,Waytowich2016,drwuTHMS2017,drwuTFS2016,drwuTNSRE2016,drwuSMC2015,drwuSMC2014}. TL leverages relevant data or knowledge from other subjects or tasks to reduce the calibration effort for a new subject or task. Traditionally, EEG signal processing (e.g., CSP filtering) and classification (e.g., TL) are performed sequentially and independently. However, recent research has shown that TL may be used to directly enhance CSP for better filtering performance \cite{Kang2009,Dalhoumi2014,Lotte2010}.

This paper focuses on TL enhanced CSPs. Its main contributions are:
\begin{enumerate}
\item We group existing TL enhanced CSPs into two categories and give a comprehensive review of them. To our knowledge, this is the first review in this direction.
\item We propose a novel TL enhanced CSP approach, and demonstrate its performance against existing approaches on EEG-based motor imagery classification.
\end{enumerate}

The rest of this paper is organized as follows: Section~\ref{sect:bk} introduces CSP and TL, and gives an overview of existing approaches for incorporating TL into CSP. Section~\ref{sect:TLCSP} proposes a new instance-based TL approach to enhance CSP. Section~\ref{sect:comp} compares the performance of all these approaches. Finally, Section~\ref{sect:conclusions} draws conclusions and points out several future research directions.

\section{Existing TL Enhanced CSP Filters} \label{sect:bk}

This section briefly introduces CSP and TL, and reviews three existing approaches for integrating them.

\subsection{Common Spatial Pattern (CSP)}

Let $X\in \mathbb{R}^{C\times T}$ be an EEG epoch, where $C$ is the number of channels and $T$ the number of time samples. For simplicity, only binary classification is considered in this paper.

CSP \cite{drwuSF2018,Blankertz2008,Ramoser2000} separates a multivariate signal into additive subcomponents which have maximum differences in variance between the two classes. Specifically, CSP finds a filter matrix to maximize the variance for one class while minimizing it for the other:
\begin{align}
W_0=\arg \max_{W} \frac{\mathrm{tr}(W^T\bar{\Sigma}_0W)}{\mathrm{tr}(W^T \bar{\Sigma}_1W)} \label{1}
\end{align}
where $W_0\in \mathbb{R}^{C\times F}$ is the filter matrix consisting of $F$ filters, $\mathrm{tr}(\cdot)$ is the trace of a matrix, $\bar{\Sigma_0}$ and $ \bar{\Sigma_1}$ are the mean covariance matrices of epochs in Classes 0 and 1, respectively. The solution $W_0$ is the concatenation of the \textit{F} eigenvectors associated with the \textit{F} largest eigenvalues of the matrix $\bar{\Sigma}_1^{-1}\bar{\Sigma}_0$.

In practice, we often construct a CSP filter matrix $W_{\ast}=[W_0, W_1]\in \mathbb{R}^{C\times 2F}$, where
\begin{align}
W_1=\arg \max_W \frac{\mathrm{tr}(W^T\bar{\Sigma}_1W)}{\mathrm{tr}(W^T \bar{\Sigma}_0W)}\label{2}
\end{align}
i.e., $W_1$ maximizes the variance for Class 1 while minimizing it for Class 0. Similar to $W_0$, $W_1$ is the concatenation of the $F$ eigenvectors associated with the $F$ largest eigenvalues of the matrix $\bar{\Sigma}_0^{-1}\bar{\Sigma}_1$. Since $\bar{\Sigma}_1^{-1}\bar{\Sigma}_0$ and $\bar{\Sigma}_0^{-1}\bar{\Sigma}_1$ have the same eigenvectors, and the eigenvalues of $\bar{\Sigma}_1^{-1}\bar{\Sigma}_0$ are the inverses of the eigenvalues of $\bar{\Sigma}_0^{-1}\bar{\Sigma}_1$, $W_1$ actually consists of the $F$ eigenvectors associated with the $F$ smallest eigenvalues of the matrix $\bar{\Sigma}_1^{-1}\bar{\Sigma}_0$. So, only one eigen-decomposition of the matrix $\bar{\Sigma}_1^{-1}\bar{\Sigma}_0$ (or $\bar{\Sigma}_0^{-1}\bar{\Sigma}_1$) is needed in computing $W_{\ast}$.

Once $W_{\ast}$ is obtained, CSP projects an EEG epoch $X\in \mathbb{R}^{C\times T}$ to $X'\in \mathbb{R}^{2F\times T}$ by:
\begin{align}
 X'=W_{\ast}^TX \label{eq:X'}
\end{align}
Usually $2F<C$, so CSP can increase the signal-to-noise ratio and reduce the dimensionality simultaneously.

After CSP filtering, the logarithmic variance feature vector is then calculated as \cite{Dalhoumi2014}:
\begin{align}
\mathbf{x}=\log\left(\frac{\mathrm{diag}(X'X'^T)}{\mathrm{tr}(X'X'^T)}\right) \label{eq:x}
\end{align}
where $\mathrm{diag}(\cdot)$ returns the diagonal elements of a matrix. $\mathbf{x}$ can be used as the input to a classifier, e.g., linear discriminant analysis (LDA).

\subsection{Transfer Learning (TL)}

TL has been extensively used in BCIs to reduce their calibration effort \cite{Jayaram2016,Waytowich2016,drwuTHMS2017,drwuTFS2016,drwuTNSRE2016}. Some basic concepts of TL are introduced in this subsection.

A \emph{domain} \cite{Pan2010,Long2014} $\mathcal{D}$ in TL consists of a feature space $\mathcal{X}$ and a marginal probability distribution $P(\mathbf{x})$, i.e., $\mathcal{D}=\{\mathcal{X},P(\mathbf{x})\}$, where $\mathbf{x}\in \mathcal{X}$. Two domains $\mathcal{D}_s$ and $\mathcal{D}_t$ are different if $\mathcal{X}_s\neq \mathcal{X}_t$, and/or $P_s(\mathbf{x})\neq P_t(\mathbf{x})$.

A \emph{task} \cite{Pan2010,Long2014} $\mathcal{T}$ in TL consists of a label space $\mathcal{Y}$ and a conditional probability distribution $Q(y|\mathbf{x})$. Two tasks $\mathcal{T}_s$ and $\mathcal{T}_t$ are different if $\mathcal{Y}_s\neq \mathcal{Y}_t$, or $Q_s(y|\mathbf{x})\neq Q_t(y|\mathbf{x})$.

Given a \emph{source domain} $\mathcal{D}_s$ with $n$ labeled samples, and a \emph{target domain} $\mathcal{D}_t$ with $m_l$ labeled samples and $m_u$ unlabeled samples, TL learns a target prediction function $f: \mathbf{x} \mapsto y$ with low expected error on $\mathcal{D}_t$, under the assumptions $\mathcal{X}_s\neq\mathcal{X}_t$, $\mathcal{Y}_s\neq\mathcal{Y}_t$, $P_s(\mathbf{x})\neq P_t(\mathbf{x})$, and/or $Q_s(y|\mathbf{x})\neq Q_t(y|\mathbf{x})$.

For example, in EEG-based motor imagery classification studied in this paper, a source domain consists of EEG epochs from an existing subject, and the target domain consists of EEG epochs from a new subject. When there are $Z$ source domains $\{\mathcal{D}_s^z\}_{z=1,...,Z}$, we can perform TL for each of them separately and then aggregate the $Z$ classifiers, or treat the combination of the $Z$ source domains as a single source domain.

\subsection{Incorporating TL into CSP: Covariance Matrix-Based Approaches} \label{sect:CM}

Since covariance matrices are used in CSP, whereas the target domain does not have enough labeled samples to reliably estimate them, a direction to incorporate TL into CSP is to utilize the source domain covariance matrices to enhance the estimation of the target domain ones.

Kang et al. \cite{Kang2009} proposed a subject-to-subject transfer approach, which emphasizes the covariance matrices of source subjects who are more similar to the target subject. They computed the dissimilarity between the target subject and each source subject by Kullback-Leibler (KL) divergence between their data distributions, and then used the inverses of these dissimilarities as weights to combine the source domain covariance matrices.

Let $p_s^z$ be the EEG data distribution in the $z$th source domain $\mathcal{D}_s^z$, which is assumed to be $C$-dimensional Gaussian with zero mean and covariance matrix $\Sigma_s^z$, i.e., $p_s^z\sim N(\mathbf{0},\Sigma_s^z)$. Let $p_t$ be the data distribution in the target domain $\mathcal{D}_t$, which is $C$-dimensional Gaussian with zero mean and covariance matrix $\Sigma_t$, i.e., $p_t\sim N(\mathbf{0},\Sigma_t)$. The KL divergence between $p_s^z$ and $p_t$ is computed as \cite{Kang2009}:
\begin{align}
KL(p_s^z,p_t)=\frac{1}{2}\left\{\log\left(\frac{|\Sigma_t|}{|\Sigma_s^z|}\right)
+\mathrm{tr}[\Sigma_t^{-1}\Sigma_s^z]-C\right\},\quad z=1,...,Z \label{8}
\end{align}
where $|\cdot|$ is the matrix determinant.

Then, the TL-enhanced covariance matrix for the target subject is computed as:
\begin{align}
\widetilde{\Sigma}_t=(1-\lambda)\Sigma_t+\lambda\sum_{z=1}^Z\alpha_z\Sigma_s^z \label{10}
\end{align}
where $\lambda$ is an adjustable parameter to balance the information from the target subject and source subjects, and
\begin{align}
 \alpha_z=\frac{1}{\gamma}\cdot\frac{1}{KL(p_s^z,p_t)} \label{9}
\end{align}
in which $\gamma=\sum_{z=1}^Z\frac{1}{KL(p_s^z,p_t)}$ is a normalization factor.

Lotte and Guan \cite{Lotte2010} proposed a similar approach for incorporating TL into CSP, based on the covariance matrices:
\begin{align}
\widetilde{\Sigma}_t=(1-\lambda)\Sigma_t+\frac{\lambda}{|S_t(\Omega)|}
\sum_{z\in S_t(\Omega)}\Sigma_s^z \label{eq:Lotte}
\end{align}
where $\Omega$ is the set of subjects whose data have been recorded previously, $S_t(\Omega)$ is a subset of subjects from $\Omega$, $|S_t(\Omega)|$ is the number of subjects in $S_t(\Omega)$, and $\lambda\in[0,1]$ is defined by
\begin{align}
   \lambda=\begin{cases}
   1,&{targetAcc\leq randAcc}\\
    0,&{targetAcc\geq selectedAcc}\\
    \frac{selectedAcc-targetAcc}{1-randAcc},&\mbox{otherwise} \label{eq:lambda}
    \end{cases}
\end{align}
in which $targetAcc$ is the leave-one-out validation accuracy on the target domain labeled samples when the classifier is trained by using only the target domain labeled samples, $selectedAcc$ is the accuracy on the target domain labeled samples when the classifier is trained by using only the labeled samples from the selected source subjects in $S_t(\Omega)$, and $randAcc$ is the classification accuracy at the chance level (e.g., 50\% for binary classification). The algorithm for determining $S_t(\Omega)$ can be found in \cite{Lotte2010}.

\subsection{Incorporating TL into CSP: A Model-Based Approach} \label{sect:MA}

Instead of learning a single set of CSP filters by aggregating information from the target subject and all (or a subset of) source subjects, as introduced in the previous subsection, Dalhoumi et al. \cite{Dalhoumi2014} proposed an approach to design a set of CSP filters for each source subject, train a classifier for each source subject according to the extracted features, and then aggregate all these source classifiers to obtain the target classifier.

Let $W^z$ and $f^z$ be the CSP filter matrix and classifier trained for the $z$th source subject, respectively, and $\{(X_j, y_j)\}_{j=1,...,m}$ be the labeled target domain data. We first filter each $X_j$ by $W^z$, extract the corresponding feature vector $\mathbf{x}_j^z$ using (\ref{eq:x}), and then feed $\mathbf{x}_j^z$ into model $f^z$ to obtain its classification $f^z(\mathbf{x}_j^z)$. The final classifier is:
\begin{align}
f(\mathbf{x})=\sum_{z=1}^Z w^zf^z(\mathbf{x}) \label{14}
\end{align}
where the weights $\mathbf{w}_*=(w^1,...,w^Z)$ are determined by solving the following constrained minimization problem:
\begin{align}
&\mathbf{w}_*=\arg\min_{\mathbf{w}}\sum_{j=1}^m\ell\left(\sum_{z=1}^Zw^zf^z(\mathbf{x}_j^z),y_j\right)\\
s.t. \quad & \sum_{z=1}^Zw^z=1 \nonumber\\
&w^z\ge 0, \ z=1,...,Z \nonumber
\end{align}
where $\ell\left(\sum_{z=1}^Zw^zf^z(\mathbf{x}_j^z),y_j\right)$ is the loss between $\sum_{z=1}^Zw^zf^z(\mathbf{x}_j^z)$ and $y_j$.

Dalhoumi et al. \cite{Dalhoumi2014} also constructed another CSP filter matrix and the corresponding classifier using the target domain data only, and compared its leave-one-out validation performance with that of $f(\mathbf{x})$ to determine which one should be used as the preferred classifier. Because the goal of this paper is to compare different TL enhanced CSP approaches, we always use $f(\mathbf{x})$.

\section{Incorporating TL into CSP: Instance-Based Approaches} \label{sect:TLCSP}

This section introduces our proposed approach for incorporating TL into CSP. It's an instance-based approach, meaning that the source domain labeled samples are combined with the target domain labeled samples in a certain way to design the CSP.

The simplest instance-base approach is to directly combine the labeled samples from the target domain and all source domains. However, this is usually not optimal because it completely ignores the individual difference: some source domain samples may be more similar to the target domain samples, so they should be given more consideration.

So, a better approach is to re-weight the source domain samples according to their similarity to the target domain samples, and then use them in the CSP. The main problem is how to optimally re-weight the source samples. We adopt the approach proposed by Huang et al. \cite{Huang2006}, which is a generic method for correcting sample collection bias and has not been used for CSP and BCIs. It assigns different weights to the source domain samples to minimize the Maximum Mean Discrepancy \cite{Belkin2006} between the source and target domains after mapping onto a reproducing kernel Hilbert space. More specifically, it solves the following constrained minimization problem:
\begin{align}
 &\min\limits_{\boldsymbol{\beta}}\left\|\frac{1}{n}\sum_{j=1}^n\beta_j\phi(\mathbf{x}_s^j)
 -\frac{1}{m}\sum_{j=1}^m\phi(\mathbf{x}_t^j) \right\|_H^2 \label{eq:MMD}\\
 s.t. \quad & 0\le\beta_j\le b, \quad j=1,...,n \nonumber\\
 &\left|\sum_{j=1}^n\beta_j-n\right|\le n\epsilon \nonumber
\end{align}
where $\mathbf{x}_s^j$ is the $j$th source domain sample, $\mathbf{x}_t^j$ is the $j$th target domain sample, $\phi(\mathbf{x})$ is a feature mapping onto a reproducing kernel Hilbert space $H$, $\boldsymbol{\beta}=(\beta_1,...,\beta_n)$ is the weight vector for the source domain samples, $n$ is the number of source domain samples, $m$ is the number of target domain samples, and $b$ and $\epsilon$ are adjustable parameters.

The source domain samples are then re-weighted by $\boldsymbol{\beta}$ and combined with the target domain samples to design a CSP filter matrix.

\section{Experiment and Results} \label{sect:comp}

This section presents a comparative study of the above TL-enhanced CSP algorithms.

\subsection{Dataset and Preprocessing}

We used Dataset 2a from BCI competition IV\footnote{\href{http://www.bbci.de/competition/iv/}{http://www.bbci.de/competition/iv/}.}, which consists of EEG data from 9 subjects. Every subject was instructed to perform four different motor imagery tasks, namely the imagination of movement of the left hand, right hand, both feet, and tongue. A training session and a test session were recorded on different days for each subject and each session is comprised of 288 epochs (72 for each of the four classes). The signals were recorded using 22 EEG channels and 3 EOG channels at 250Hz and bandpass filtered between 0.5Hz and 100Hz.

Only the 22 EEG channels were used in our study. We further processed them using the Matlab EEGLAB toolbox \cite{Delorme2004}. They were first down-sampled to 125Hz. Next a bandpass filter of 8-30 Hz was applied as movement imagination is known to suppress idle rhythms in this frequency band contra-laterally \cite{Pfurtscheller2006}. As we consider binary classification in this paper, only EEG signals corresponding to the left and right hand motor imageries were used. More specifically, EEG epochs between 1.5 and 3.5 seconds after the appearance of left or right hand motor imagery cues were used.

\subsection{Algorithms}

We compared the performance of the following seven CSP algorithms:
\begin{enumerate}
\item \emph{Baseline 1 (BL1)}, which uses only the small amount of target domain labeled samples to design the CSP filters and the LDA classifier, and applies them to target domain unlabeled samples. That's, BL1 does not use any source domain samples.
\item \emph{Baseline 2 (BL2)}: which combines all source domain samples to design the CSP filters and the LDA classifier, and applies them to target domain unlabeled samples. That's, BL2 does not use any target domain labeled samples.
\item \emph{Baseline 3 (BL3)}, which directly combines all source domain samples and target domain labeled samples, designs the CSP filters and the LDA classifier, and applies them to target domain unlabeled samples.
\item \emph{Covariance matrix-based approach 1 (CM1)}, which is the approach proposed by Kang et al. \cite{Kang2009}, as introduced in Section~\ref{sect:CM}. $\lambda=0.5$ was used in our study.
\item \emph{Covariance matrix-based approach 2 (CM2)}, which is the approach proposed by Lotte and Guan \cite{Lotte2010}, as introduced in Section~\ref{sect:CM}.
\item \emph{Model-based approach (MA)}, which is the approach introduced in Section~\ref{sect:MA}.
\item \emph{Instance-based approach (IA)}, which is our proposed algorithm: it first solves the constrained optimization problem in (\ref{eq:MMD}) for the weights of the source domain samples, then combines target domain labeled samples and the weighted source domain samples to train CSP filters and the LDA classifier, and next applies them to target domain unlabeled samples.
\end{enumerate}

There were 9 subjects in our dataset. Each time we picked one as our target subject, and the remaining 8 as the source subjects. For the target subject, we randomly reserved 40 epochs (20 epochs per class) as the training data pool, and used the remaining 104 epochs as our test data. We started with zero target domain training data, trained different CSP filters using the above 7 algorithms, and evaluated their performances on the test dataset. We then sequentially added 2 labeled epochs (1 labeled epoch per class) from the reserved training data pool to the target domain training dataset till all 40 epochs were added. Each time we trained different CSP filters using the above 7 algorithms and evaluated their performances on the test dataset. We repeated this process 30 times to obtain statistically meaningful results.

\subsection{Results}

The performances of the 7 algorithms are shown in Fig.~1, where the first 9 subfigures show the performances on the individual subjects. Observe that some subjects, e.g., Subjects 2 and 5, were more difficult to deal with than others, and there was no approach that always outperformed others; however, when $m$, the number of target domain labeled epochs, was small, our proposed algorithm (IA) achieved the best performance for 5 out of the 9 subjects.

The last subfigure of Fig.~1 shows the average performance across the 9 subjects. Observe that:
\begin{enumerate}
\item When $m$ was small, all other methods outperformed BL1. Particularly, when $m=0$, BL1 cannot build a model because it used only subject-specific calibration data, but all other algorithms can, because they can use data from the source subjects. This suggests that all TL-enhanced CSP algorithms are advantageous when the target domain has very limited labeled epochs.
\item BL2 outperformed BL1 and BL3 when $m$ was small, but as $m$ increased, all other algorithms outperformed BL2. This suggests that there is large individual difference among the subjects, so incorporating target domain samples is necessary and beneficial.
\item Generally, all TL-enhanced CSP algorithms outperformed the three baselines, suggesting the effectiveness of TL. Particularly, our proposed algorithm (IA) achieved the best performance when $m$ was small. This is favorable, as we always want to achieve the best calibration performance with the smallest number of subject-specific calibration samples.
\end{enumerate}

\begin{figure}\centering
\includegraphics[width=\linewidth]{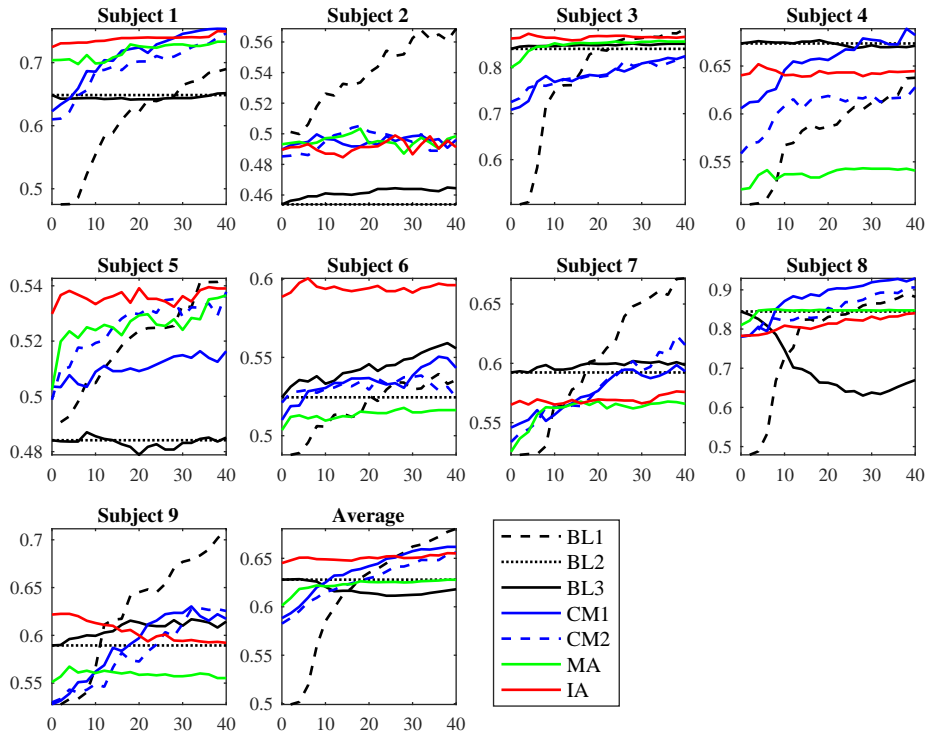}
\caption{Classification accuracies of the 7 CSP approaches, when the number of target domain labeled samples increases.} \label{fig:1}
\end{figure}

\section{Conclusions} \label{sect:conclusions}

CSP is a popular spatial filtering approach to increase the signal-to-noise ratio of EEG signals. However, it is a supervised approach, which needs some subject-specific calibration data to design. This is time-consuming and not user-friendly. A promising approach for shortening or even completely eliminating this calibration session is TL, which leverages relevant data or knowledge from other subjects or tasks. This paper reviewed three existing approaches for incorporating TL into CSP, and also proposed a new TL enhanced CSP approach. Experiments on motor imagery classification demonstrated the effectiveness of these approaches. Particularly, our proposed approach achieved the best performance when the number of target domain calibration epochs is small.

The following directions will be considered in our future research:
\begin{enumerate}
\item Use the Riemannian mean instead of the Euclidean mean in estimating the mean class covariance matrices in CSP \cite{Barachant2010}. As the covariance matrix of each epoch is semi-positive definite, they are located on a Riemannian manifold instead of in an Euclidean space. So, the Riemannian means may be more reasonable than the Euclidean means in CSP.
\item Use also TL enhanced classifiers, e.g., weighted domain adaptation \cite{drwuTHMS2017,drwuTNSRE2016}.
\item Extend the TL enhanced CSPs from classification to regression, using a fuzzy set based approach similar to the one proposed in \cite{drwuSF2018}.
\end{enumerate}


\begin{thebibliography}{10}
\providecommand{\url}[1]{\texttt{#1}}
\providecommand{\urlprefix}{URL }

\bibitem{Barachant2010}
Barachant, A., Bonnet, S., Congedo, M., Jutten, C.: Common spatial pattern
  revisited by {R}iemannian geometry. In: {IEEE} Int'l Workshop on Multimedia
  Signal Processing. pp. 472--476. London, UK (October 2010)

\bibitem{Belkin2006}
Belkin, M., Niyogi, P., Sindhwani, V.: Manifold regularization: A geometric
  framework for learning from labeled and unlabeled examples. Journal of
  Machine Learning Research  7,  2399--2434 (2006)

\bibitem{Blankertz2008}
Blankertz, B., Tomioka, R., Lemm, S., Kawanabe, M., Muller, K.R.: Optimizing
  spatial filters for robust {EEG} single-trial analysis. {IEEE} Signal
  Processing Magazine  25(1),  41--56 (2008)

\bibitem{Dalhoumi2014}
Dalhoumi, S., Dray, G., Montmain, J.: Knowledge transfer for reducing
  calibration time in brain-computer interfacing. In: Proc. 26th {IEEE} Int'l
  Conf. on Tools with Artificial Intelligence. Limassol, Cyprus (November 2014)

\bibitem{Delorme2004}
Delorme, A., Makeig, S.: {EEGLAB}: an open source toolbox for analysis of
  single-trial {EEG} dynamics including independent component analysis. Journal
  of Neuroscience Methods  134,  9--21 (2004)

\bibitem{Erp2012}
van Erp, J., Lotte, F., Tangermann, M.: Brain-computer interfaces: Beyond
  medical applications. Computer  45(4),  26--34 (2012)

\bibitem{Huang2006}
Huang, J., Smola, A.J., Gretton, A., Borgwardt, K.M., Scholkopf, B.: Correcting
  sample selection bias by unlabeled data. In: Proc. Int'l. Conf. on Neural
  Information Processing Systems. pp. 601--608. Vancouver, Canada (December
  2006)

\bibitem{Jayaram2016}
Jayaram, V., Alamgir, M., Altun, Y., Scholkopf, B., Grosse-Wentrup, M.:
  Transfer learning in brain-computer interfaces. {IEEE} Computational
  Intelligence Magazine  11(1),  20--31 (2016)

\bibitem{Kang2009}
Kang, H., Nam, Y., Choi, S.: Composite common spatial pattern for
  subject-to-subject transfer. Signal Processing Letters  16(8),  683--686
  (2009)

\bibitem{Lance2012}
Lance, B.J., Kerick, S.E., Ries, A.J., Oie, K.S., McDowell, K.: Brain-computer
  interface technologies in the coming decades. Proc. of the {IEEE}  100(3),
  1585--1599 (2012)

\bibitem{Long2014}
Long, M., Wang, J., Ding, G., Pan, S.J., Yu, P.S.: Adaptation regularization: A
  general framework for transfer learning. {IEEE} Trans. on Knowledge and Data
  Engineering  26(5),  1076--1089 (2014)

\bibitem{Lotte2010}
Lotte, F., Guan, C.: Learning from other subjects helps reducing brain-computer
  interface calibration time. In: Proc. {IEEE} Int'l. Conf. on Acoustics Speech
  and Signal Processing ({ICASSP}). Dallas, TX (March 2010)

\bibitem{Makeig2012}
Makeig, S., Kothe, C., Mullen, T., Bigdely-Shamlo, N., Zhang, Z.,
  Kreutz-Delgado, K.: Evolving signal processing for brain-computer interfaces.
  Proc. of the {IEEE}  100(Special Centennial Issue),  1567--1584 (2012)

\bibitem{Nicolas-Alonso2012}
Nicolas-Alonso, L.F., Gomez-Gil, J.: Brain computer interfaces, a review.
  Sensors  12(2),  1211--1279 (2012)

\bibitem{Pan2010}
Pan, S.J., Yang, Q.: A survey on transfer learning. {IEEE} Trans. on Knowledge
  and Data Engineering  22(10),  1345--1359 (2010)

\bibitem{Pfurtscheller2006}
Pfurtscheller, G., Brunner, C., Schlogl, A., da~Silva, F.L.: Mu rhythm
  (de)synchronization and {EEG} single-trial classification of different motor
  imagery tasks. NeuroImage  31(1),  153--159 (2006)

\bibitem{Ramoser2000}
Ramoser, H., Muller-Gerking, J., Pfurtscheller, G.: Optimal spatial filtering
  of single trial {EEG} during imagined hand movement. {IEEE} Trans. on
  Rehabilitation Engineering  8(4),  441--446 (2000)

\bibitem{Waytowich2016}
Waytowich, N.R., Lawhern, V.J., Bohannon, A.W., Ball, K.R., Lance, B.J.:
  Spectral transfer learning using {I}nformation {G}eometry for a
  user-independent brain-computer interface. Frontiers in Neuroscience  10,
  430 (2016)

\bibitem{Wolpaw2002}
Wolpaw, J.R., Birbaumer, N., McFarland, D.J., Pfurtscheller, G., Vaughan, T.M.:
  Brain-computer interfaces for communication and control. Clinical
  Neurophysiology  113(6),  767--791 (2002)

\bibitem{drwuTHMS2017}
Wu, D.: Online and offline domain adaptation for reducing {BCI} calibration
  effort. {IEEE} Trans. on Human-Machine Systems  47(4),  550--563 (2017)

\bibitem{drwuSF2018}
Wu, D., King, J.T., Chuang, C.H., Lin, C.T., Jung, T.P.: Spatial filtering for
  {EEG}-based regression problems in brain-computer interface ({BCI}). {IEEE}
  Trans. on Fuzzy Systems  26(2),  771--781 (2018)

\bibitem{drwuSMC2014}
Wu, D., Lance, B.J., Lawhern, V.J.: Transfer learning and active transfer
  learning for reducing calibration data in single-trial classification of
  visually-evoked potentials. In: Proc. {IEEE} Int'l Conf. on Systems, Man, and
  Cybernetics. San Diego, CA (October 2014)

\bibitem{drwuTNSRE2016}
Wu, D., Lawhern, V.J., Hairston, W.D., Lance, B.J.: Switching {EEG} headsets
  made easy: {Reducing} offline calibration effort using active weighted
  adaptation regularization. {IEEE} Trans. on Neural Systems and Rehabilitation
  Engineering  24(11),  1125--1137 (2016)

\bibitem{drwuSMC2015}
Wu, D., Lawhern, V.J., Lance, B.J.: Reducing offline {BCI} calibration effort
  using weighted adaptation regularization with source domain selection. In:
  Proc. {IEEE} Int'l Conf. on Systems, Man and Cybernetics. Hong Kong (October
  2015)

\bibitem{drwuRG2017}
Wu, D., Lawhern, V.J., Lance, B.J., Gordon, S., Jung, T.P., Lin, C.T.:
  {EEG}-based user reaction time estimation using {R}iemannian geometry
  features. {IEEE} Trans. on Neural Systems and Rehabilitation Engineering
  25(11),  2157--2168 (2017)

\end{thebibliography}
\end{document}